\documentclass[conference]{IEEEtran}
\IEEEoverridecommandlockouts

\usepackage[T1]{fontenc}     

\usepackage{booktabs}

\usepackage{cite}
\usepackage{amsmath,amssymb,amsfonts}
\usepackage{algorithmic}
\usepackage{graphicx}
\usepackage{textcomp}
\usepackage{xcolor}
\usepackage[hyphens]{url}  
\def\BibTeX{{\rm B\kern-.05em{\sc i\kern-.025em b}\kern-.08em
    T\kern-.1667em\lower.7ex\hbox{E}\kern-.125emX}}
\begin{document}

\title{Designing for Self-Regulation in Informal Programming Learning: Insights from a Storytelling-Centric Approach
	
{
	\footnotesize \textsuperscript{}}
\thanks{}

}

\author{\IEEEauthorblockN{Sami Saeed Alghamdi}
\IEEEauthorblockA{\textit{Open Lab} \\
\textit{Newcastle University}\\
Newcastle upon Tyne, UK \\
s.s.a.alghamdi2@newcastle.ac.uk}
\and
\IEEEauthorblockN{Christopher Bull}
\IEEEauthorblockA{\textit{Open Lab} \\
\textit{Newcastle University}\\
Newcastle upon Tyne, UK\\
christopher.bull@newcastle.ac.uk}
\and
\IEEEauthorblockN{Ahmed Kharrufa}
\IEEEauthorblockA{\textit{Open Lab} \\
\textit{Newcastle University}\\
Newcastle upon Tyne, UK \\
ahmed.kharrufa@newcastle.ac.uk}
}

\maketitle

\begin{abstract}
Many people learn programming independently from online resources and often report struggles in achieving their personal learning goals. Learners frequently describe their experiences as isolating and frustrating, challenged by abundant uncertainties, information overload, and distraction, compounded by limited guidance. At the same time, social media serves as a personal space where many engage in diverse self-regulation practices, including help-seeking, using external memory aids (e.g., self-notes), self-reflection, emotion regulation, and self-motivation. For instance, learners often mark achievements and set milestones through their posts.

In response, we developed a system consisting of a web platform and browser extensions to support self-regulation online. The design aims to add learner-defined structure to otherwise unstructured experiences and bring meaning to curation and reflection activities by translating them into learning stories with AI-generated feedback. We position storytelling as an integrative approach to design that connects resource curation, reflective and sensemaking practice, and narrative practices learners already use across social platforms.

We recruited 15 informal programming learners who are regular social media users to engage with the system in a self-paced manner; participation concluded upon submitting a learning story and survey. We used three quantitative scales and a qualitative survey to examine users’ characteristics and perceptions of the system’s support for their self-regulation. User feedback suggests the system’s viability as a self-regulation aid. Learners particularly valued in-situ reflection, automated story feedback, and video annotation, while other features received mixed views. We highlight perceived benefits, friction points, and design opportunities for future AI-augmented self-regulation tools.
\end{abstract}

\section{Introduction}
Many people interested in programming take a self-directed approach to learning, drawing on a wide range of informal online resources ( e.g., \cite{b1,b2,b3,b4}). According to a 2024 Stack Overflow survey, programming learners engage more frequently with open-ended, nonlinear materials such as forums, tutorials, technical documentation, and social media platforms (e.g., YouTube, Twitch, and X) than with textbooks or structured e-learning courses (i.e., MOOCs) \cite{b5}. However, this inclination toward informal online learning is often accompanied by limited support for synthesizing information (i.e., actively combining insights from multiple resources) or engaging in sustained reflection. As a result, much of their learning—whether intentional or incidental—tends to be fragmented and transient \cite{b2,b4,b6,b7}. 


While such sporadic engagement can be productive in the short term (e.g., solving a specific task-based problem), it often lacks mechanisms for deeper understanding and misses opportunities to consolidate learning—both of which are important for constructive learning in programming \cite{b8}, especially given its exploratory nature \cite{b1,b4}. To regain structure and persist in such learning, learners adapt social media to support self-regulation strategies such as publicly reflecting on progress, maintaining motivation through shared achievements, validating their learning approach with peers, and regulating emotions \cite{b6,b9,b10}.

Despite the prevalence of these self-regulatory practices in social media spaces, there has been little focused inquiry into learners’ experiences in these settings (e.g.\cite{b2,b6,b11,b12}). Most research on programming learning and self-regulation focuses on formal settings such as university classrooms and MOOCs, where learners benefit from structured curricula, instructor guidance, and feedback (e.g., \cite{b13,b14,b15,b16,b17,b18}). In contrast, informal learners operate in loosely organized, socially driven contexts that require dedicated support. Addressing this challenge calls for moving beyond frameworks developed for instructor-led education toward design approaches grounded in the lived practices of self-directed learners \cite{b9,b19,b20,b21}.

However, recognizing these needs is not the same as responding to them. While some prior work has examined informal learning needs \cite{b2,b6} and explored learners’ use of online platforms \cite{b3,b4,b12,b22,b23}, few studies have translated these insights into tool designs, and even fewer have evaluated them with informal learners. Thus, it remains unclear how to create tools that support self-regulation while fitting seamlessly into learners’ everyday online routines. To address this gap, we explored storytelling as a promising design element for supporting self-regulation in programming (e.g., \cite{b24,b25,b26,b27,b28,b29}). 

Transforming learning experiences into narratives promotes externalization and fosters reflection, which helps surface hidden obstacles and strengthen metacognitive awareness \cite{b6,b30}. Stories also serve as engaging and memorable formats for sharing practical knowledge and experiences, which are prevalent in technical domains \cite{b31}. Prior research has shown that learners use storytelling as a form of social currency within online communities \cite{b6,b32}, enabling learners to access peer networks, identify with others’ experiences, and sustain motivation—key elements of self-regulated learning \cite{b1,b33}. Thus, we position storytelling as an integrative approach to design that connects informal practices (resource curation, reflective and sensemaking, and narrative practices \cite{b3,b6,b11,b34,b35}) that learners already employ during learning explorations and across social platforms.

To explore this storytelling-centric approach in a working system, we implemented a system comprising a backend web platform (Link removed for anonymity) and three browser extensions (Link removed for anonymity). The design supports learners in organizing learning into Resources, Tags, and Stories, enabling them to reflect on online materials, tag them to construct learning paths across multiple resources, and generate narrative summaries of their progress, complemented by AI-generated feedback on their informal learning activities.

 This paper makes two contributions. First, we report the design rationale and implementation of storytelling-centric tools to support self-regulation in informal programming learning, grounded in prior findings about learner needs and practices. Second, we present findings from a short-term study of informal learners’ experience using the system, highlighting perceived benefits, areas of friction, and design opportunities for future AI-augmented self-regulation tools.
 
This study investigates the following research questions:
\begin{itemize}
	\item RQ1: How can storytelling be integrated into the design of self-regulation support for informal programming learners? 
    
	\item RQ2: What are learners’ initial perceptions of the usefulness and usability of storytelling-centric self-regulation tools?
    \item 	RQ3: What design considerations can be derived from learners’ feedback?
\end{itemize}
The remainder of the paper is structured as follows. We first review related work on the design space of tools supporting self-regulation of learning. We then present the system’s design and its rationale, followed by our study methodology and key findings. We conclude with a discussion of implications for designing self-regulation tools.

\section{Related Works}
We position our work within tool-based approaches supporting self-regulation among informal programming learners. This section reviews prior literature on informal programming learning, outlines key dimensions of self-regulation, surveys existing self-regulation tools, and explores the potential of storytelling to support self-regulation in learning to program.

\subsection{Informal Programming Learning and Self-Regulation}
 Informal programming learners benefit from self-regulation support \cite{b1,b36,b37}. They require assistance in pursuing self-learning independently outside formal and structured environments. Such settings, including university courses and MOOCs, frequently report high dropout rates among programming learners \cite{b38,b39,b40,b41}. Self-regulation refers to organizing and utilizing personal and environmental resources effectively to support learning goals. It can be fostered through cognitive skills, like applying learning strategies, and metacognitive skills, such as planning, monitoring, and evaluation. Motivation is also key for promoting self-regulation \cite{b1,b42,b43,b44}. Given the broad range of components, it is essential to prioritize the self-regulatory needs of specific populations to design effective solutions that address their needs.
 
To understand the nature of informal learning, Gao et al. \cite{b3} proposed a model (COIL) that describes how informal learners (i.e., developers learning new APIs) engage in cycles of content curation, organization, and integration to codebase, offering a lens to understand the informal learning experience. The COIL model is based on studying how developers practice API learning, and such a model can be extended to cover how accompanied self-regulation is practiced during such learning for other learners such as less experienced, casual or novices learners \cite{b45}. 

Relatedly, Lie et al. \cite{b34} frame such informal learning activity as daily online sensemaking developers engage on web browsers. They introduce Crystalline, a tool that automatically gathers and structures information into tables as developers explore and navigate the web to alleviate the manual curation burden and support decision-making during learning. 
For developers, such sensemaking processes can support self-regulation by reducing cognitive load. However, novice learners may require additional support beyond sensemaking, such as externalizing thoughts and offloading intentions and future goals. \cite{b46,b47,b48}. Notably, these offloading mechanisms are not facilitated by structuring information for sensemaking alone. Therefore, balancing learner agency with automation represents a critical design tension when the goal is to support the direction of learning.

In terms of social need, Chelana et al. identified “conversational programmers” as informal programming learners who do not aim to integrate into a codebase through their learning \cite{b12,b49}. Instead, they focus on other purposes, such as enhancing communication with programming teams. Alghamdi et al. further synthesized prior literature on informal programming learning, surfacing peer support and self-efficacy as important factors to support this population \cite{b1}. These social needs are actively facilitated through social media platforms \cite{b6,b50,b51,b52,b53}. The aforementioned studies highlight the social and motivational aspects of programming learning, stressing the importance of interventions that foster relatedness \cite{b54} to support learning regulation and maintain continuity.

\subsection{Tool Support for Self-Regulation of Learning}
 Self-regulated learning literature in formal programming education is well-established and growing(e.g., \cite{b43,b55,b56,b57,b58,b59}). It emphasizes the teacher's role in teaching strategies, designing pedagogical interventions, and creating assignments to improve course outcomes. Correspondingly, tools in this area are primarily designed for structured environments, such as extensions to learning management systems or MOOCs, assuming an instructor-led environment with prepared materials and predefined learning goals. Further, many tools are neither tailored to programming nor designed for informal learning; prior literature reviews list tools that essentially assume structured, formal education settings \cite{b18,b60}. As such, programming learners operating in informal settings lack dedicated support for self-regulation and often rely on non-specialized tools, including social platforms, to manage their learning \cite{b1,b61}.
 
Having a clear learning goal is essential for informal learning regulation and progression \cite{b6,b36}. However, informal learners, especially novices, often lack this clarity \cite{b2,b6}, overwhelmed by abundant resources and uncertainty \cite{b1}. This represents a key early obstacle in learning to program independently. Thus, Rimika et al. \cite{b11,b62} designed paper-based mock-ups to investigate the needs of individuals in technical domains for self-monitoring their learning. The study examined interactive visual self-monitoring tools for informal learners and found that such tools can enhance learners’ ability to reflect on past experiences and plan future goals, especially when they offer user control, such as filterable progress views.

Beyond goal setting, informal learners must also navigate the challenges of maintaining focus in distracting digital environments. Self-control in web browsing is an important topic in online experience literature for general users. It aims to reduce negative online experiences, such as distractions, information overload, and compulsive use \cite{b63}. Various tools such as blocking websites, usage logs and screen timers aim to manage browsing experience \cite{b63,b64,b65}. However, these tools primarily focus on behavioural control and underserve other self‑regulation needs, such as metacognitive and motivational ones, which are essential for learning, warranting future research for an integrative approach.

Similarly, commercially available tools such as Forest \cite{b66} and Habitica \cite{b67} aim to help users ‘stay focused’ and ‘achieve goals’ through gamified experiences. However, their role in informal programming learning remains unclear. These tools do not seamlessly integrate with informal learning practices that involve curation, sensemaking, and self-directed exploration \cite{b3,b34}. Self-regulation of learning involves processes that promote ongoing engagement with learning goals, utilizing reflective practices and social learning \cite{b6,b52,b68,b69}.

\subsection{Storytelling and Programming}
 The potential of storytelling is recognized in computing and software engineering education and practice (e.g., \cite{b25,b28,b29,b30,b31,b70,b71,b72}), primarily to enhance instructional delivery or serve as a documentation strategy that improves code comprehension. Digital storytelling was discussed as a promising approach for fostering participation and collaboration in multidisciplinary software teams \cite{b72}. Further, Wuilmart et al. \cite{b28}, state that stories can help developers make sense of complex codebases by linking narratives to specific development tasks. Nonetheless, developers frequently report difficulty initiating narrative \cite{b28}, underscoring the need for storytelling scaffolds. 
 
In the same vein, Amber et al. explored developers' various needs for sensemaking tools \cite{b73}. For instance, they designed “Sodalite” \cite{b74} and “Adamite” \cite{b75}, which address developers' need to author ‘stories’ about their code as a long-form of documentation and to engage with API documentation via annotation. They note that such tools enable them to overcome the issue of poor documentation, allowing self-notes and explanations for their sensemaking or sharing with other developers. Furthermore, Kery et al. \cite{b30} focused on literate programming tools, such as Jupyter notebooks. They demonstrated that data scientists organize cells in notebooks, forming narratives that help them understand their coding task explorations. 

While these efforts highlight storytelling's value in formal education and developers' contexts (e.g., engaging teaching, better code documentation, API learnability, and sensemaking during exploration), they lack focus on specific issues for novices learning informal programming. Namely, challenges that precede coding practice including high uncertainty, lack of goal clarity, and lack of communication efficiency in seeking community support \cite{b1,b2}. In response, Alghamdi et al. analysed learner-authored narratives across various programming channels on social media platforms \cite{b6}; their findings show that learners often narrate their experiences in public posts to document progress, process setbacks, and connect with peer communities, to self-regulate their learning. 

Further, three levels of tool support are identified and proposed to scaffold storytelling \cite{b6}, as follows: \textbf{(1) interaction with resources}  (i.e., curation in alignment with \cite{b3,b34}), \textbf{(2) managing experiences} (i.e., classification, subgoal labeling or tagging, in alignment with \cite{b40,b76,b77}), and \textbf{(3) externalization of these experiences}, utilising Generative AI abilities (Aligning with these proposals \cite{b6,b74}). Together, these levels of support aim to assist learners in articulating and managing their learning experiences. As such, storytelling surfaced as a promising component to facilitate self-regulation and was proposed as a design element grounded in the learners' existing experiences and needs \cite{b6}. Nevertheless, current literature has not yet extended to demonstrate how to position storytelling in a design for informal learners or investigated its practical utility for self-regulation. 
 
\subsection{Synthesis}
 While the potential of storytelling within programming education and practice has been examined, its implementation as a learner-authored and self-regulation support element remains largely underexplored. Informal programming learners usually rely on curation \cite{b3,b34} and sensemaking of fragmented online resources. Also, they depend on social media to self-regulate during informal activities \cite{b6,b10}. Key aspects of self-regulation for this group highlight the importance of social connections and reflection on experiences \cite{b1,b36}, which align with principles of self-determination \cite{b54} and experiential learning theory \cite{b78}. Storytelling has been identified as a promising method to fulfil both needs \cite{b6}. However, current tools do not adequately address the specific self-regulation dimensions identified for informal programming learners. Therefore, this area warrants further research and design to enhance independent, self-directed learning.  

\section{Design}
In this section, we will illustrate a scenario of a novice informal learner, drawing inspiration from previous literature and fieldwork that investigated aspects related to informal programming learning ( e.g., \cite{b1,b2,b3,b4,b6,b36,b45}). Then, we outline the design requirements for tools to support self-regulation in such informal experiences, and present the resulting tool design that responds to these requirements.

\subsection{Motivating Example}

Consider Amy, a self-taught novice programmer eager to “get serious” about coding. Her initial enthusiasm soon gives way to uncertainty: Should she explore JavaScript frameworks like React or Vue? Focus on Python and data science? Try backend development with Java, or game development with Rust?

Without clear goals, guidance from peers, or visible role models, Amy cycles through tutorials without direction. She tentatively chooses Python, only to encounter new decisions—Should she use notebooks or local files? Build GUIs with Tkinter or PyQt? Eventually, she turns to Vue.js, drawn by its HTML-like familiarity, but soon encounters friction with tooling such as NPM.

Though Amy engages with a wide array of technologies, her learning remains fragmented. She relies on ChatGPT for code generation but struggles to understand how and why the solutions work. Motivated moments—such as building a desktop app—fade when she cannot retrace her steps or make sense of past decisions. Throughout her journey, Amy lacks continuity, self-efficacy, and peer support. Most importantly, she lacks a goal that personally resonates with her to maintain momentum. 

Her experience illustrates common barriers faced by informal learners: cognitive overload, resource fragmentation, limited feedback, and declining motivation. Without tools that scaffold decision-making, support resource organization, and render learning goals visible and personally meaningful, many spikes in motivation fade away easily, and exploration efforts are lost and cannot be built upon. 

\subsection{Design Rationale}
The previous scenario highlights several challenges in informal programming learning: resuming learning after interruptions, making sense of fragmented progress, and lacking goal clarity. As discussed in related works, most existing solutions can support specific self-regulation components (e.g., summaries to reduce cognitive load or self-control to manage distraction alone). However, they might impose rigid workflows or fail to support seamless transitions between exploration and reflection; in other words, they fail to integrate seamlessly into the informal experience identified (e.g., learners curating, organizing, integrating into codebases, or narrating progress on social media). 

In response, we position storytelling as an \textbf{integrative approach to design} that connects resource curation, reflective and sensemaking practice, and narrative practices learners already employ across social platforms. By scaffolding storytelling, learners may reconnect with their exploratory learning over time, solidifying their progress toward clearer goals and holding the potential to address self-regulation needs for informal learners. 

\subsection{Requirements}
We formulated the following design requirements, informed by the motivating scenario, design rationale and related literature:
 \begin{itemize}
    \item \textbf{R1. In-situ reflection during curation} that allows learners to reflect across diverse resources and intentions during the act of curation, supporting later sensemaking.
    
    \item \textbf{R2. Resource organization} to help learners transform fragmented content into coherent learning paths, so that past efforts can be built upon.
    
    \item \textbf{R3. Turning curated learning paths into stories} to support post-reflection and to enable the following:
    \begin{enumerate}
        \item \textbf{R3.1. Social mechanisms for community engagement} via sharing and exporting to social platforms.
        \item \textbf{R3.2. AI-generated feedback} based on learners’ stories, derived from curation and reflections, to support post-reflection, helping with goal refinement or adjustment.
    \end{enumerate}
\end{itemize}

\subsection{System Structure}
\begin{figure*}[!t]
  \centering
  \includegraphics[width=\textwidth]{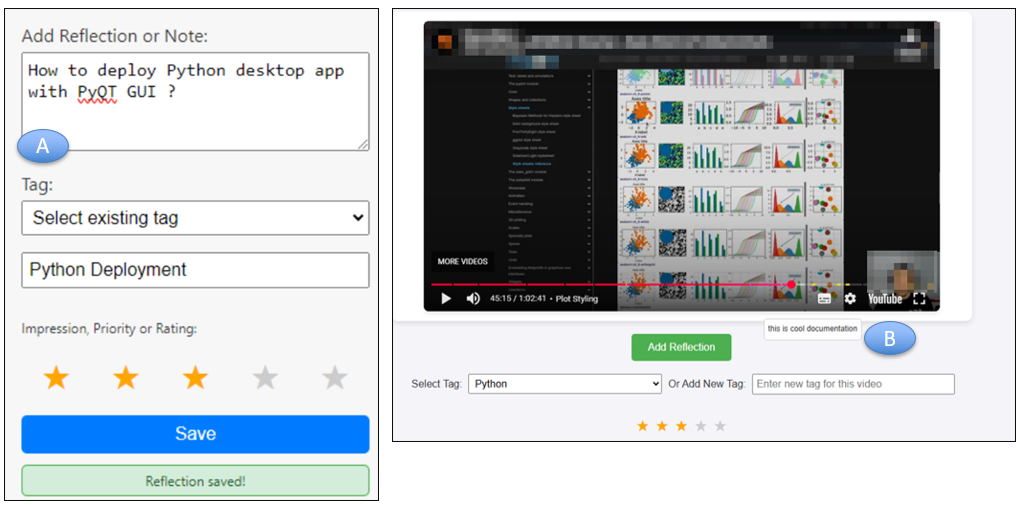}
  \caption{Story Curator and YouTube Annotator extensions. (A) The Story Curator extension allows learners to tag, rate, and reflect on online resources, helping with organization and immediate offloading.  (B) The YouTube Annotator extension allows timestamped reflections while watching videos, displaying reflections as balloons. Clicking a balloon jumps to that part of the video.  These tools enable in situ reflection during exploratory learning (R1, R2). Tagged reflections can be converted into stories, each tag representing a new story.}
  \label{fig1}
\end{figure*}

The system was developed as a minimum viable product (MVP), comprising a platform and three browser extensions to assist learners in externalizing their learning experiences. The tools provide support ranging from the curation stage to sharing: \textbf{(1) Story Curator} allows for tagging, rating, and reflecting on web resources; \textbf{(2) YouTube Annotator} that enables learners to create time-linked annotations as videos play, as well as tag and rate the reflections on videos. Both extensions focus on enhancing the curation of resources and facilitating reflections during exploratory learning (see Fig. \ref{fig1}). 

\textbf{ (3) The backend platform}, mainly provides technical infrastructure for API management, but it also provides a \textbf{visual representation} of the system logic that organizes learning experience into \textit{Resources, Tags and Stories}, (see Fig.2). Learners can generate stories of their learning experiences, as shown in Fig 3. \textbf{(4) Learner Eye} allows learners to monitor their recent activities (namely, the period since they last added resources and reflections, the last time new tags or learning paths were initiated, and the last story they created). It provides ways to share versions of stories modified for various social media. It also offers features to control their social media use (See Fig. 4).

  \begin{figure}[!t]
  \centering
  \includegraphics[width=\columnwidth]{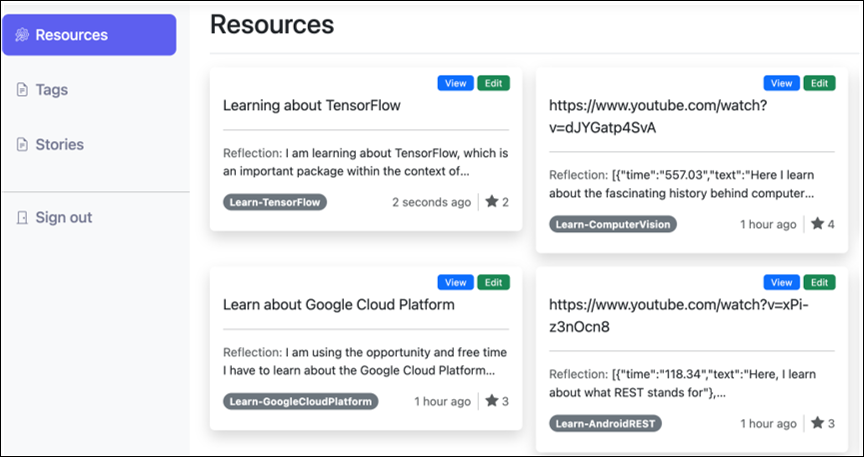} 
  \caption{The platform primarily stores data and manages API connections between front-end extensions and OpenAI. It also displays resources from the curation process and allows for story creation through a dedicated tab in the left sidebar.
}
  \label{fig:overview}
\end{figure}

 \begin{figure}[!t]
  \centering
  \includegraphics[width=\columnwidth]{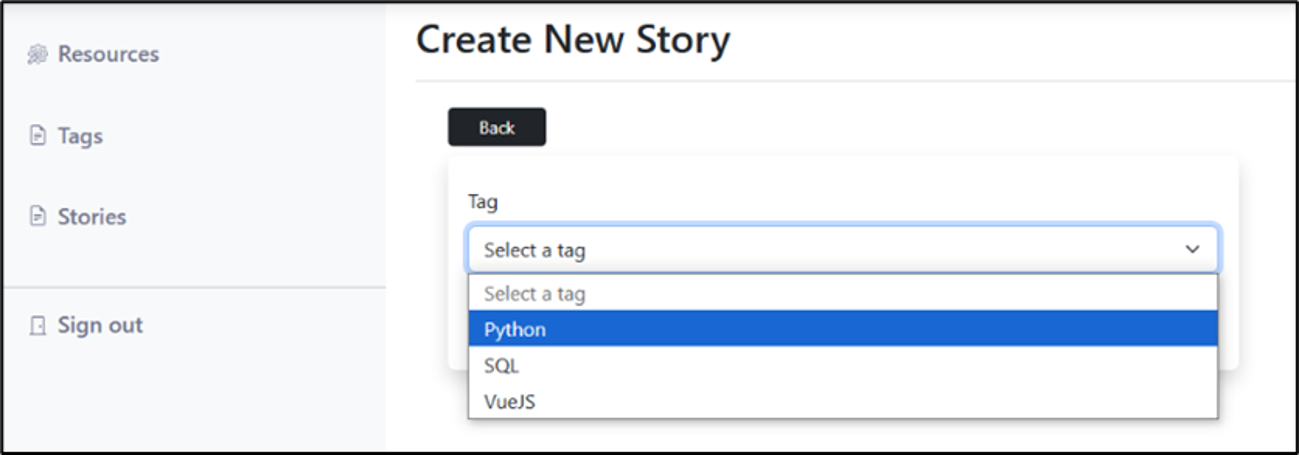} 
  \caption{Learners can generate stories of their learning experiences via the stories tab in the left sidebar. The shown menu lists the learners’ created tags (or learning paths). Each tag represents a curation process, stored reflections, and links to resources, such as anchors to multiple YouTube segments they previously curated within a tag. }
  \label{fig:overview}
\end{figure}

 \begin{figure}[!t]
  \centering
  \includegraphics[width=\columnwidth]{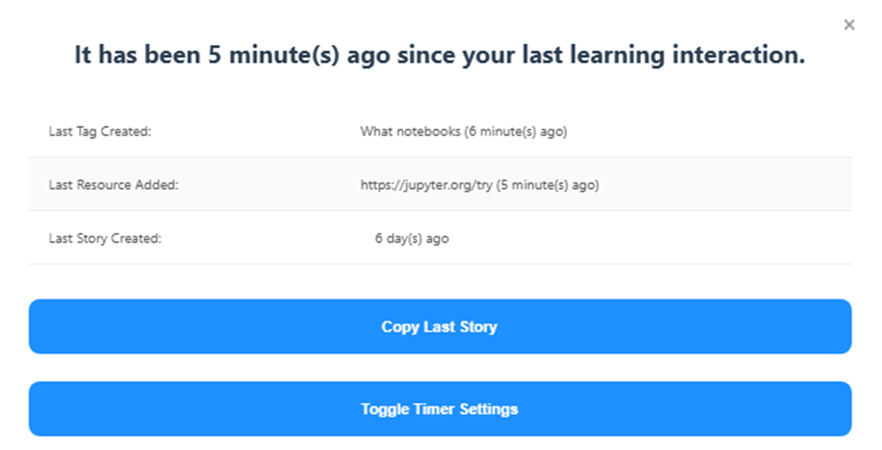} 
  \caption{Learner Eye extension aims to help learners monitor their own curation and storytelling activity; it also provides a mechanism to share stories on social media and control excessive use on those platforms. The screenshot shows a pop-up when learners visit social media platforms, depicting recent activity.
}
  \label{fig:overview}
\end{figure}
 \begin{figure}[!t]
  \centering
  \includegraphics[width=\columnwidth]{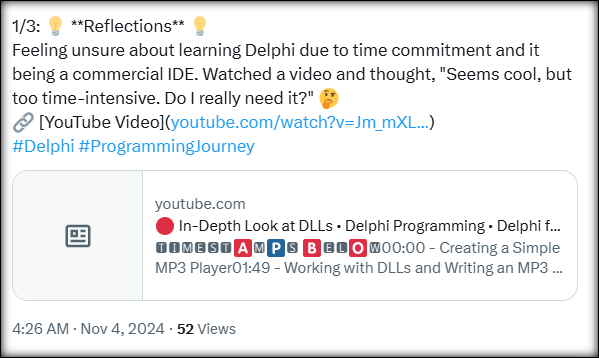} 
  \caption{A learning story shared to X using Learner Eye. The author made sure to protect the associated X profile for anonymity
}
  \label{fig:overview}
\end{figure}

\subsection{Storytelling Mechanism}
Users can generate stories after the tagged curation process to make sense of their experience and get AI feedback. Their reflections via curation tools (Fig. 1) are sent with a Prompt to the OpenAI API. The platform stores the generated stories for use on the platform and by Learner Eye extension (Fig. 4). 
The extension re-generates a version of the last story for the context of a social media platform. For example, if the user is on X (previously Twitter), it will modify the last learning story created in the platform as a series of numbered posts, or a ‘Thread’, copied to the clipboard (see Fig.4 and Fig.5).
It is noted that story quality depends on the curation process. 

The general structure of a story, as stored on the platform, follows the format below:


\begin{quote}
\ttfamily

Title of a story\\

 \ \ \ Listing of users' reflections \\

 \ \ \ Keywords for overall reflections \\

AI-Feedback



\end{quote}
To achieve this structure, the prompt was iteratively refined to generate a learner-authored or first-person story, avoiding superficial or general advice (e.g., referring to searching for a solution on Stack Overflow).

\begin{figure}[!t]
  \centering
  \includegraphics[width=\columnwidth]{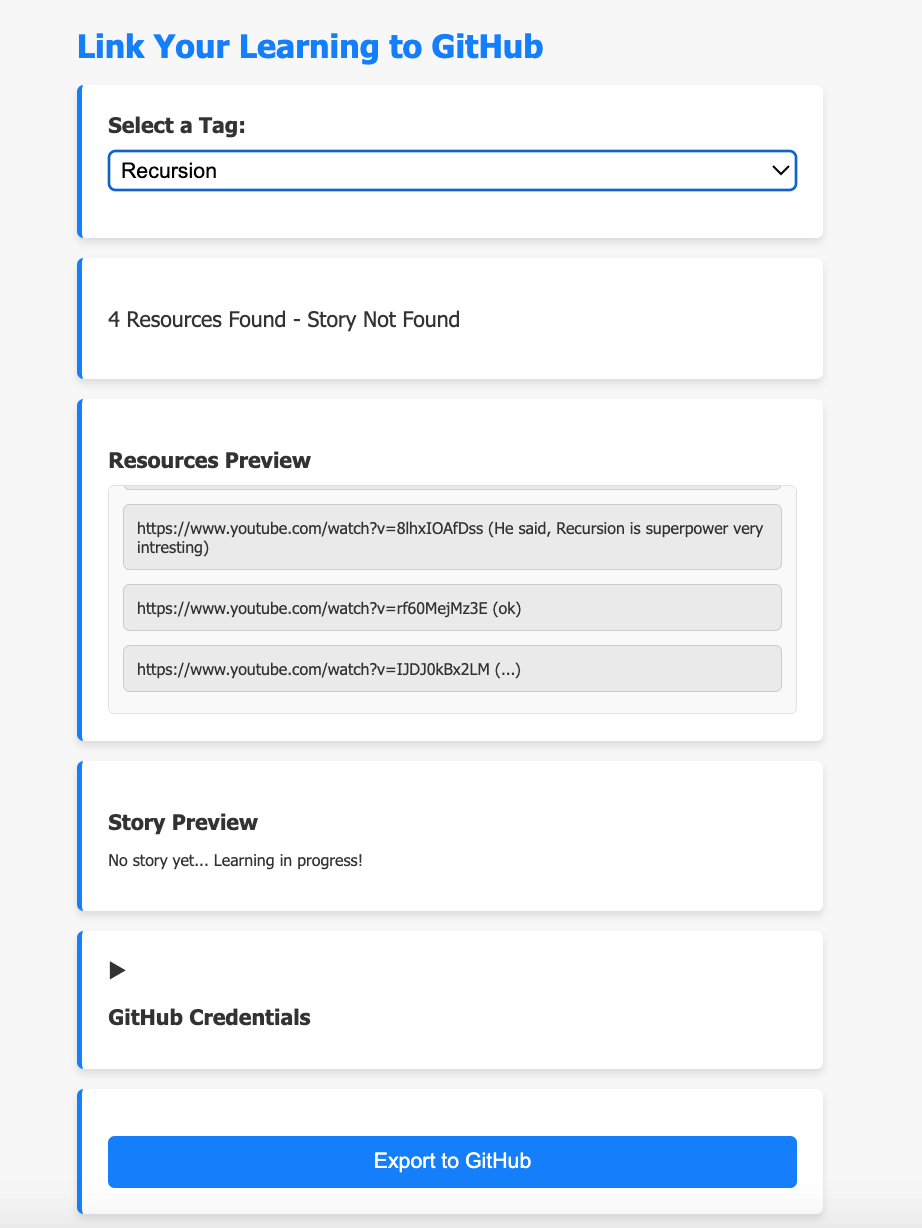} 
  \caption{Options page of Story Curator allow exporting to GitHub. Learners can export learning to GitHub, where reflection and resources are organised as MD files. The story, if generated, will be in the readme file. If not yet created, the number of resources and their times will be added, noting that learning is in progress. 
 }
 
  \label{fig:overview}
\end{figure}

  \begin{figure}[!t]
  \centering
  \includegraphics[width=\columnwidth]{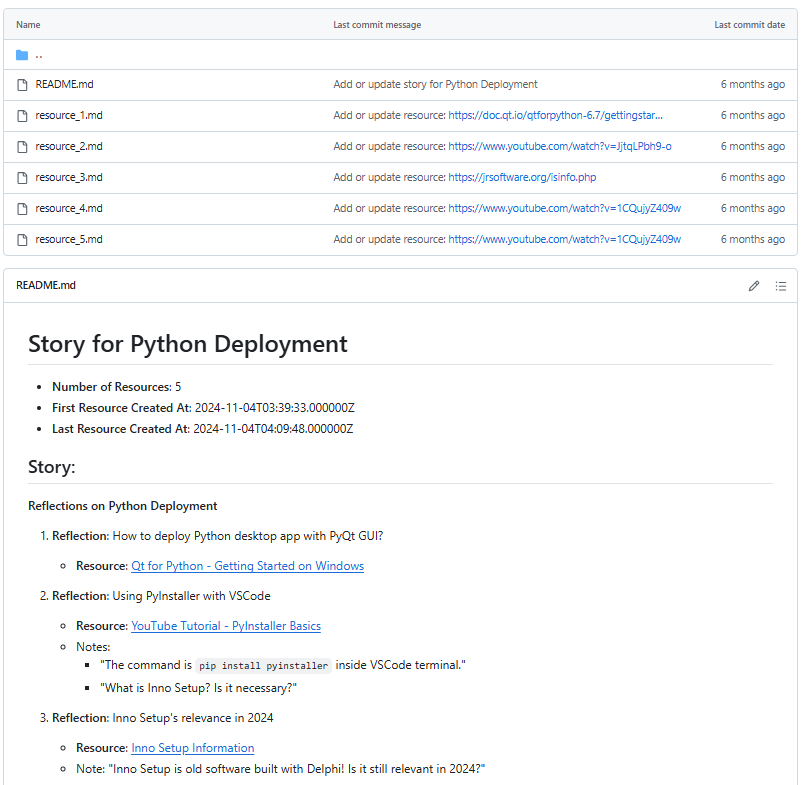} 
  \caption{Screenshot of exported learning to GitHub. Author made sure to make the repository private for anonymity.
}
  \label{fig:overview}
\end{figure}

\subsection{Bridging Fragmented Learning via Tool-Mediated Workflows: From Curation to Sharing}
The curation tools aligns with learners’ practices and preferences for informal exploration. The tools support them during this curation phase by offloading their intentions, questions, and notes. They can also rate resources for later evaluation. Tags allow them to initiate learning paths, keeping track of simulationous learning experiences that occasionally interleave. To manage experiences, the tools offer a mechanism to monitor current activity to pick up where the last curation left off immediately.

Further, the system can render a radar chart via the Learner Eye options page, showing the number of resources on each tag. Allowing learners to evaluate their overall learning. The platform enables learners to storytell their learning experiences with generated feedback. Then Learner Eye extension allow learners to share modified versions of stories for social media, while also providing tools to manage the excessive social media use and nudge for learning. 

The system allows learners to effortlessly export selected learning experiences to GitHub (See Fig. 6 and Fig. 7 ). The repository name is the tag, the README file contains the story, and the curation processes are organised as MD files with learner reflections timestamped. This feature nudges practice by giving a uniform structure and streamlining and scaffolding the transition to practice, tackling an issue reported by learners in prior research.

\section{Method}
Our approach aligns with the Research-through-Design methodology \cite{zimmerman2007}: we developed a system to explore the "wicked" challenge \cite{wickedproblem} of supporting self-regulation in informal programming learning. Such challenges lack a single optimal solution and instead demand context-sensitive responses that can vary in effectiveness. As part of this approach, we focused on a specific population: programming learners who are regular social media users, to reflect the design problem's contextual nature and ground our intervention in actual learner practices.

All participants were self-reported programming learners who regularly use social media in their everyday lives. Three quantitative scales and a qualitative survey were used to examine users’ characteristics, perceptions of the system’s overall usability, and feedback on specific features in relation to perceived aid to their self-regulation. We detail the study procedure in this section.

\subsection{Participant Recruitment}
We recruited 15 informal programming learners through Prolific \cite{prolific} (n = 14) and social media platforms (n = 1). Initially, we attempted to recruit participants via social media via a recruiting survey; although several expressed interest (n=42), only one ultimately responded to follow-up emails. All 15 participants self-identified as programming learners who regularly use social media. Their programming experience ranged from novices to more experienced learners and professionals. All participants were compensated fairly for their time.

\subsection{Procedure}
Participants were guided through the tool installation via email instructions , blog posts and three surveys: recruitment, guidance, and evaluation. They were asked to use the three web extensions, Story Curator, YouTube Annotator, and Learner Eye, alongside the leading platform (Link removed). They were encouraged to \textbf{(1)} tag and reflect on programming learning resources of their choice, \textbf{(2)} generate learning stories about their learning experiences, and \textbf{(3)} explore other features (e.g., sharing stories and viewing visualizations). They were asked to generate at least one story that combined a minimum of three resources, along with reflections on those resources.

 The leading researcher reviewed each survey submission to verify the inclusion of a genuine story based on curated resources, serving as minimal evidence of engagement with the system. This process was supported by Google Analytics to check the number of actual participants. Unsatisfactory submissions (i.e., submitting a survey without pasting a story) were returned to Prolific for revision and to include a proper story based on curation and engagement with the system's features.
 
\subsection{Instruments Rationale}
 After the trial, participants completed a quantitative and qualitative survey. The quantitative component focused on participant characteristics and overall system usability, using validated scales: the General Self-Efficacy Scale (8 items) \cite{b79}, the Sense of Coherence scale (13 items) \cite{b82,b83} (SoC), and the User Experience Scale Short version (8 items) \cite{b84} (UEQ-S).  The qualitative component explored participants’ perceptions of the system’s key features and its perceived impact on their self-regulation. 

We used the validated self-efficacy and SoC due to challenges recruiting learners through social media; many completed the surveys but did not respond to follow-up emails. This highlighted how costly and difficult it can be to sustain participant engagement remotely. We included these scales to characterise our sample better and support future research and replication. While the sample size is small, we hope this provides a more in-depth description of our participants. We elaborate on these scales as follow: 
 \subsubsection{Sense of Coherence}
  The Sense of Coherence Scale measures three constructs: how individuals perceive life as comprehensible, manageable, and meaningful\cite{b80}. A higher sense of coherence score is associated with better stress management and overall well-being. Research indicates that individuals with high SoC perceive challenges as navigable, whereas those with lower SoC are more likely to adopt avoidant strategies , such as procrastination \cite{b81}.
  
  \subsubsection{Self-efficacy}
  Self-efficacy refers to an individual’s belief in their capability to achieve goals \cite{lippke2020}. Prior research shows that learners with high self-efficacy tend to approach tasks with greater persistence. In contrast, those with low self-efficacy often doubt their skills, avoid tasks perceived as risky, and are more likely to disengage\cite{pajares1996}.
  
These two scales may reveal variation and potential correlations in learners’ approaches to self-directed programming tasks within our tools. However, given the small sample size, we do not treat such findings as reliable. Nor do we claim that learners with specific psychological characteristics will consistently respond to the design in a particular way. Instead, the scales offer quantitative documentation of our sample, support the interpretation of participant feedback, and provide a foundation for future research.

\subsection{Analyses}
We used descriptive statistics to summarize results from the SoC, Self Efficacy, and UEQ-S scales, calculating mean (M) and standard deviation (SD) to characterize the sample and support replication. Qualitative responses to open-ended questions were analysed using inductive thematic analysis to identify recurring themes in learners’ experiences and feedback. 

\section{Results}
We divide our findings into two main sections. The first section describes user characteristics to help replicate our results and offer insights for future research, especially concerning low engagement in system trials, which can be challenging and costly. The second section presents users' feedback on the system and their opinions on its effectiveness in supporting self-regulation.
\subsection{Users’ Characteristics}
Fifteen informal programming learners (7 females) participated in the study. Their self-reported programming experience ranged from novice to advanced. The average age was 30.7 years (SD = 8.05), with a range from 19 to 46.

On the self-efficacy scale (5-point), participants averaged 3.88 (SD = 0.49), suggesting relatively close or similar perceptions of self-efficacy across the sample.
In contrast, scores on the SoC (maximum score: 91) were more dispersed (M = 53.42, SD = 8.35, range: 43 to 72), reflecting broader individual differences in sense of coherence.
\subsection{System Feedback}
\subsubsection{Overall User Experience}

  \begin{figure}[!t]
  \centering
  \includegraphics[width=\columnwidth]{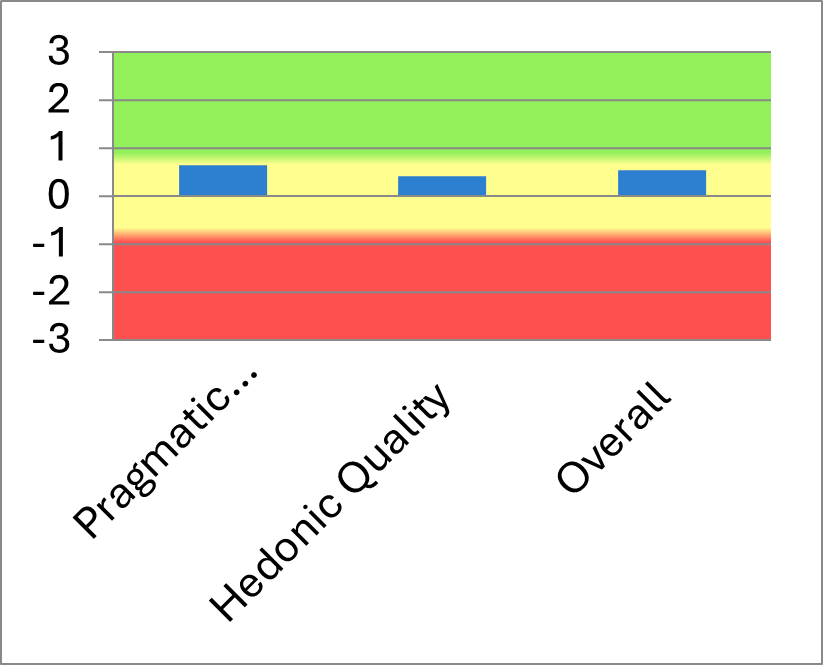} 
  \caption{ The UEQ-S scale shows a slightly above average positive user experience for the overall system experience. 
}
  \label{fig:overview}
\end{figure}

To assess whether the system is usable overall, participants completed the UEQ-S, where scores range from -3 (horribly bad) to +3 (extremely good). The average score was 0.54 (SD = 1.29), indicating slightly positive experiences, especially given that the system is an unpolished MVP, with wide variability (range: -1.75 to 3).
\subsubsection{User Feedback on System Elements}
To evaluate participants’ perceptions of specific system features, we administered 17 5-points agreement-scale items assessing the extent to which these features aligned with the intended design goals. Selected survey items are presented below:

\begin{itemize}
  \item “The \textbf{system} helps me reflect effectively on my learning experiences”.
  \item “The \textbf{story} generation feature enables me to summarize and build on my learning”.
  \item “The ability to \textbf{share stories} on social media enhances my motivation and engagement”.
  \item “The \textbf{tagging} feature helps me organize and navigate my learning pathways”.
  \item “The \textbf{web extensions} are easy to use and integrate into my daily learning routine”.
  \item “The \textbf{story} helps identify actionable next steps”.
\end{itemize}

 For the seven items assessing overall system features, the mean rating was 3.35 (SD = 0.58). The five items related to the browser extensions yielded a slightly higher mean of 3.47 (SD = 0.70). 
 
 The storytelling feature (five items) received the highest agreement among the components evaluated, with a mean of 3.73 (SD = 0.78) across five items.



\subsubsection{Perceived Support for Self-Regulation }
We used a 3-point scale on six items to investigate the perceived impact of the system on participants' self-regulation in informal learning (See Fig. 9). This simplified scale was chosen to reduce response fatigue, particularly following the preceding multi-scale instruments, and to prompt clearer, more decisive judgments about impact.

Participants’ responses across all items were generally positive, notably,  no one selecting “disagree” for any statement. Eleven participants agreed that the system helps them easily resume previous learning sessions, while four were neutral.

Thirteen participants agreed that it provides helpful context or reminders to pick up where they left off, with two responding neutrally. Twelve participants agreed that the system encourages reflection on earlier learning goals, and the same number agreed that it helps refine goals based on progress and challenges; three participants were neutral on each of these items.
 
Fewer participants agreed that the system supports adapting learning strategies over time, only six indicated agreement, while nine were neutral. A similar distribution was observed for the item about organizing or structuring learning activities, where nine participants agreed and six responded neutrally.

  \begin{figure}[!t]
  \centering
  \includegraphics[width=\columnwidth]{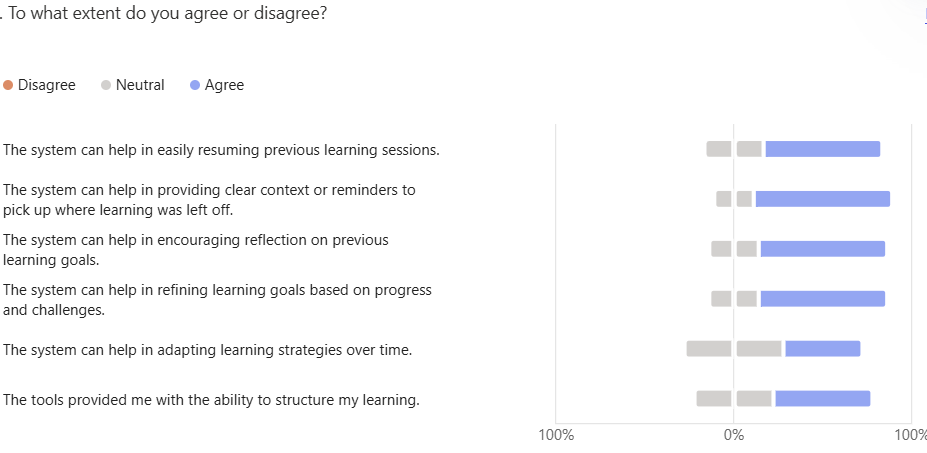} 
  \caption{ Six items focused on perceived support for self-regulation. 
}
  \label{fig:overview}
\end{figure}

\subsection{Qualitative Responses}
Our survey included several open-ended questions. Examples of open-ended questions used:

\begin{itemize}
    \item “Think about your past learning experiences without the help of tools. Now, tell us: how confident do you feel about achieving your learning goals with the help of such a system?”
    \item “Do the connections between tags, resources, and stories feel natural and intuitive?”
    \item “Do you think the storytelling approach can effectively guide you toward achieving your learning goals?”
    \item “What did you like most about the system and its tools?”
\end{itemize}
 We categorized the open-ended responses into three themes: positive experiences, negative experiences, and user suggestions.
\subsubsection{Positive Experiences}
 Most of the positive feedback focused on the curation affordances and the generated stories, which participants described as motivating and empowering. These features were appreciated for combining personal reflections with added insights, helping users track their learning progress and structure their efforts more effectively. The following quotes highlight several ways in which participants experienced the system as positively supporting their learning.

Learners appreciated the \textbf{clarity}: \textit{“That’s been the most useful feature for me so far. The keyword recognition from the resources and the way it breaks them down step by step into stories… it’s just very well done”.}

Others valued the \textbf{structured} nature of the experience: \textit{“The tools are not cutting-edge technologies, but at least they offer you a well structured method of learning and feedback which I found really helpful”.}

The storytelling approach was seen as adding  \textbf{meaningful} context: \textit{“... the storytelling approach can be highly effective (…) makes learning more memorable by providing relatable and meaningful contexts”.}

\textbf{Externalization} helped learners maintain motivation: \textit{“Seeing your learning journey as a narrative helps you stay connected to your goals, even during challenging times”}.

\textbf{Combining reflections with added AI-insights} was positively received: \textit{“... stories are very helpful and cool. It combines nicely and gives interesting information when click on the details”}. 

Some participants suggested that the system can help \textbf{use time more effectively}: \textit{“It would make the learning process more effective and less time consuming, which is great”}. 

A potential boost in \textbf{self-efficacy} reported: \textit{“Such a system would make it easier for me to recall what I learned earlier and refresh my memory, simplifying my learning and thus increasing my confidence in achieving my learning goals”.}

\subsubsection{Negative Experiences}
Negative responses often stemmed from a lack of engagement in real learning activities or incomplete exploration of the system’s features. However, the most critical feedback centred on the Learner Eye web extension, which integrates features to self-control social media use and handle sharing stories to social media. 

Some participants perceived them as interrupting, and they expressed \textbf{reluctance} toward social sharing. One participant explained, \textit{“I did not use the story-sharing feature; I don’t think I’ve seen it. I would not use such a feature because I am not keen on sharing my learning stories with others”} and \textit{“I have not used the feature of sharing stories on social networks like Instagram because for me that is very private and I prefer to do it myself”}. 

Further concerns were raised about self-control feature (see Fig. 4) perceived as \textbf{disruptive}: \textit{“I am not yet convinced that I would achieve my goals with its help, I was often disturbed by pop-ups when I was focused on tasks”}. Some participants also struggled with \textbf{interface} navigation. For instance, one remarked, \textit{“I would have liked the YouTube annotation if I could have figured out how to find it again once I was done with it”}. Another raised issues of \textbf{privacy}: “\textit{(…) how much information the plugins have access to, and what they can do with it”. }

\subsubsection{Suggestions}
Some participants provided suggestions, focusing on the system category, integration, and personalisation. One participant proposed a shift in \textbf{how the system is framed} and used: \textit{“I believe the system should be less of a reflection tool and more of a learning notebook, where one can make notes as they learn about what they learn, where to find materials on the topic, or make bookmarks”.} 

Another emphasized the need to \textbf{simplify the setup}: \textit{“I would prefer only 1 add-on with all options, not 3”.} Specific suggestions were also made to improve the way video annotations feed into story generation, \textit{“I think stories should handle video reflections differently; they could provide them in a list with actual links to the videos at the bookmarked moment”.} 

Further, there was interest in \textbf{adding more control to the storytelling} element: \textit{“Maybe give two or three story options so that the user can choose the one they like better”.}

\section{Discussion}
In this paper, our focus was on demonstrating storytelling as a design element within a self-regulation tool. Findings from both quantitative and qualitative data indicate the potential of this approach to support self-regulation. In this section, we reflect on the design decisions we made (addressing RQ1), highlight key findings from user feedback on the system (addressing RQ2), and conclude with design consideration for future work (addressing RQ3).

\subsection{Integrating Storytelling into Self-Regulation Tool Design (RQ1)}

By a storytelling-centric approach, we mean tools intentionally designed to externalise learning experiences. Our system organize informal programming learners’ experience around three elements, Resources, Tags, and Stories, which correspond to three levels of learner engagement: curation, organisation, and externalisation. This framing makes learning visible, evaluable, and shareable. This approach aligns with prior work discussed in the Related Work section; it explicitly extends the COIL model proposed by Gao et al. \cite{b3} and draws inspiration from the design suggestions provided by Alghamdi et al. \cite{b6}, making it suitable for describing the experience of self-regulation for informal programming learners in social media.

These elements form a conceptual design for integrating storytelling into self-regulation support: Resources enable content collection (e.g., annotating YouTube segments), Tags organise materials into evolving learning paths, and Stories synthesise activity into personal narratives. The design addresses gaps in existing tools discussed in related works, particularly the lack of support for reflection, continuity, and learner-driven synthesis.

Framing informal learning activity around curation, organisation, and externalisation offers a transferable structure for embedding storytelling into tool design to support self-regulation in informal contexts.

\subsection{Perceived Value and Friction Points (RQ2)}
Quantitative and qualitative findings showed that participants found the system usable and valuable for supporting their learning self-regulation. The integration of learners’ reflections with AI-generated feedback was appreciated. Participants described various ways in which the stories supported their learning, primarily by consolidating fragmented activities into a cohesive structure, increasing clarity, adding meaning, enhancing productivity, and saving time. Most learners found the curation process (transforming tagged resources into stories) clear and sensible.

Learners also favoured tools that support precise, in-the-moment interactions, such as in YouTube Annotation, which allows marking specific moments within a video, over tools that operate at the level of entire resources (e.g., tagging or curating whole videos or articles, as in the Story Curator). This suggests a preference for more granular, context-sensitive support during curation stage.

On the other hand, friction emerged around interface complexity (the use of multiple extensions), concerns about social sharing, and negative feedback on self-control features. Most participants preferred private use, viewing storytelling as a personal sensemaking tool more than a shared artefact.

\subsection{Considerations for Future Work (RQ3)}
In future work, we are interested in wrapping the storytelling-centric conceptual design with a more polished interface and introducing more fine-grained curation capabilities and control over generated stories. Also at the organization stage, we can add more control over tags, such as merging multiple learning paths , connect them or branch. The goal is to increase learners’ autonomy in how they interact with and organize resources, learning paths and stories. 

Additionally, we plan to move self-control and sharing features into a control panel or dashboard. Such functionality in personal learning contexts should remain optional and user-initiated. Although we offered mechanisms for control or opting out, these were initially enforced—an approach that was received negatively. Participants expressed concerns, both implicitly and explicitly, about potential violations of privacy and autonomy. This suggests that self-regulation cannot be effectively supported through imposed constraints, but rather through carefully designed, learner-driven mechanisms.

We believe there is significant room for future work to investigate how a storytelling-centric approach could inform the design of AI-augmented tools that support self-regulation—by assisting learners across key stages of informal learning (curation, organization, storytelling, and sharing), while preserving their agency and autonomy.

\subsection{Limitations}
This study is limited by a small sample size (n = 15) and a short evaluation period. Although tasks were self-paced, most participants appeared to complete the surveys with minimal effort, likely motivated by the task reward. While validated scales (SoC, Self-Efficacy, UEQ-S) may strengthen the study and offer well-rounded insights, we did not, within the scope of this paper, extend to claim any causal effects based on the feedback interpretation. A longer-term study is needed to examine potential effects on learning outcomes and explore differential impacts across learner profiles. 

Moreover, our work focused on social media users. While a vast population, we recognise that our approach may not benefit those who are not social media users or are not dependent on online resources, acknowledging those who prefer more structured and instructor-led approaches to learn to program.

\section{Conclusion}
This paper demonstrated a storytelling-centric approach to designing for self-regulation. Our findings indicate that this approach integrates well with informal learning experiences. It holds potential, enabling learners to reconnect with fragmented progress, clarify goals, and engage in sustained reflection. Participants valued the fine-grained curation of videos and perceived the storytelling element as adding clarity, structure, and meaning to their experience. On the other hand, the sharing feature was largely ignored during the trial, and some learners explicitly criticized the self-control aspects. This study contributes a conceptual design illustrating how storytelling can be integrated into a self-regulation solution for informal programming learners, highlighting features that were positively perceived as well as those that were less preferred.


\vspace{12pt}
\color{red}


\begin{thebibliography}{00}
\bibitem{b1}	S. S. Alghamdi, C. Bull, and A. Kharrufa, ‘Exploring the Support for Self-Regulation in Adult Online Informal Programming Learning: A Scoping Review’, in Proceedings of the 2023 Conference on Innovation and Technology in Computer Science Education V. 1, Turku Finland: ACM, Jun. 2023, pp. 361–367. doi: 10.1145/3587102.3588811.	
\bibitem{b2}	R. Chaudhury, P. J. Guo, and P. K. Chilana, ‘“There’s no way to keep up!”: Diverse Motivations and Challenges Faced by Informal Learners of ML’, in 2022 IEEE Symposium on Visual Languages and Human-Centric Computing (VL/HCC), Roma, Italy: IEEE, Sep. 2022, pp. 1–11. doi: 10.1109/VL/HCC53370.2022.9833100.	
\bibitem{b3}	G. Gao, F. Voichick, M. Ichinco, and C. Kelleher, ‘Exploring Programmers’ API Learning Processes: Collecting Web Resources as External Memory’, in 2020 IEEE Symposium on Visual Languages and Human-Centric Computing (VL/HCC), Dunedin, New Zealand: IEEE, Aug. 2020, pp. 1–10. doi: 10.1109/VL/HCC50065.2020.9127274.	
\bibitem{b4}	M. Beth Kery and B. A. Myers, ‘Exploring exploratory programming’, in 2017 IEEE Symposium on Visual Languages and Human-Centric Computing (VL/HCC), Raleigh, NC: IEEE, Oct. 2017, pp. 25–29. doi: 10.1109/VLHCC.2017.8103446.	

\bibitem{b5} ‘Stack Overflow Developer Survey 2024’. [Online]. Available:  \url{https://survey.stackoverflow.co/2024/}. [Accessed: 7 May 2025].


\bibitem{b6}	S. S. Alghamdi, C. Bull, and A. Kharrufa, ‘Thematic Analysis of Self-Regulation Narratives in Textual Posts by Informal Programming Learners on Social Media’, in 2024 IEEE Symposium on Visual Languages and Human-Centric Computing (VL/HCC), Liverpool, United Kingdom: IEEE, Sep. 2024, pp. 223–235. doi: \url{10.1109/VL/HCC60511.2024.00033}.	
\bibitem{b7}	M. B. Kery, B. E. John, P. O’Flaherty, A. Horvath, and B. A. Myers, ‘Towards Effective Foraging by Data Scientists to Find Past Analysis Choices’, in Proceedings of the 2019 CHI Conference on Human Factors in Computing Systems, Glasgow Scotland Uk: ACM, May 2019, pp. 1–13. doi: 10.1145/3290605.3300322.

\bibitem{b8}	N. R. Boyer, S. Langevin, and A. Gaspar, ‘Self direction \& constructivism in programming education’, in Proceedings of the 9th ACM SIGITE conference on Information technology education - SIGITE ’08, 2008. doi: 10.1145/1414558.1414585.	


\bibitem{b9}	D. Lambton-Howard, P. Olivier, V. Vlachokyriakos, H. Celina, and A. Kharrufa, ‘Unplatformed Design: A Model for Appropriating Social Media Technologies for Coordinated Participation’, in Proceedings of the 2020 CHI Conference on Human Factors in Computing Systems, 2020. doi: 10.1145/3313831.3376179.	
\bibitem{b10}	N. Dabbagh and A. Kitsantas, ‘The role of social media in self-regulated learning’, Int. J. Web Based Communities, vol. 9, no. 2, p. 256, 2013, doi: 10.1504/IJWBC.2013.053248.	



\bibitem{b11}	R. Chaudhury and P. K. Chilana, ‘Designing Visual and Interactive Self-Monitoring Interventions to Facilitate Learning: Insights from Informal Learners and Experts’, IEEE Trans. Vis. Comput. Graph., pp. 1–12, 2024, doi: 10.1109/TVCG.2024.3366469.	
\bibitem{b12}	P. K. Chilana, R. Singh, and P. J. Guo, ‘Understanding Conversational Programmers: A Perspective from the Software Industry’, in Proceedings of the 2016 CHI Conference on Human Factors in Computing Systems, San Jose California USA: ACM, May 2016, pp. 1462–1472. doi: 10.1145/2858036.2858323.	
\bibitem{b13}	D. Loksa and A. J. Ko, ‘The Role of Self-Regulation in Programming Problem Solving Process and Success’, in Proceedings of the 2016 ACM Conference on International Computing Education Research, Melbourne VIC Australia: ACM, Aug. 2016, pp. 83–91. doi: 10.1145/2960310.2960334.


\bibitem{b14}	J. Prather, B. A. Becker, M. Craig, P. Denny, D. Loksa, and L. Margulieux, ‘What Do We Think We Think We Are Doing?: Metacognition and Self-Regulation in Programming’, in ICER 2020 - Proceedings of the 2020 ACM Conference on International Computing Education Research, Association for Computing Machinery, 2020, pp. 2–13. doi: 10.1145/3372782.3406263.	
\bibitem{b15}	J. Prather et al., ‘First Things First: Providing Metacognitive Scaffolding for Interpreting Problem Prompts’, in Proceedings of the 50th ACM Technical Symposium on Computer Science Education, Minneapolis MN USA: ACM, Feb. 2019, pp. 531–537. doi:\url{10.1145/3287324.3287374}.	
\bibitem{b16}	L. Silva, A. Gomes, and A. Mendes, ‘Investigating Students’ Usage of Self-regulation of Learning Scaffoldings in a Computer-based Programming Learning Environment’, in Proceedings of the 55th ACM Technical Symposium on Computer Science Education V. 1, Portland OR USA: ACM, Mar. 2024, pp. 1244–1250. doi: 10.1145/3626252.3630885.	
\bibitem{b17}	M. Manso-Vazquez, M. Caeiro-Rodriguez, and M. Llamas-Nistal, ‘An xAPI Application Profile to Monitor Self-Regulated Learning Strategies’, IEEE Access, vol. 6, pp. 42467–42481, 2018, doi: 10.1109/ACCESS.2018.2860519.	
\bibitem{b18}	R. P. Alvarez, I. Jivet, M. Perez-Sanagustin, M. Scheffel, and K. Verbert, ‘Tools Designed to Support Self-Regulated Learning in Online Learning Environments: A Systematic Review’, IEEE Trans. Learn. Technol., vol. 15, no. 4, pp. 508–522, Aug. 2022, doi: 10.1109/TLT.2022.3193271.	
\bibitem{b19}	P. Dourish, ‘The Appropriation of Interactive Technologies: Some Lessons from Placeless Documents’, Comput. Support. Coop. Work CSCW, vol. 12, 4, pp. 465–490, 2003, doi: 10.1023/a:1026149119426.	


\bibitem{b20} F. Tosi, ‘From User-Centred Design to Human-Centred Design and the User Experience’, in *Design for Ergonomics*, A. Author, Ed., Springer Series in Design and Innovation, vol. 2. Cham, Switzerland: Springer International Publishing, 2020, pp. 47–59. [Online]. Available: https://doi.org/10.1007/978-3-030-33562-5\_3


\bibitem{b21}	A. Nguyen, J. Lämsä, A. Dwiarie, and S. Järvelä, ‘Lifelong learner needs for human-centered self-regulated learning analytics’, Inf. Learn. Sci., vol. 125, no. 1/2, pp. 68–108, Jan. 2024, doi: 10.1108/ILS-07-2023-0091.	

\bibitem{b22}	A. Alaboudi and T. D. LaToza, ‘An Exploratory Study of Live-Streamed Programming’, in 2019 IEEE Symposium on Visual Languages and Human-Centric Computing (VL/HCC), Memphis, TN, USA: IEEE, Oct. 2019, pp. 5–13. doi: 10.1109/VLHCC.2019.8818832.	


\bibitem{b23}	M. B. Garcia, I. C. Juanatas, and R. A. Juanatas, ‘TikTok as a Knowledge Source for Programming Learners: a New Form of Nanolearning?’, IEEE Xplore, pp. 219–223, 2022, doi:\url{ 10.1109/ICIET55102.2022.9779004}.	
\bibitem{b24}	E. Caminotti and J. Gray, ‘The effectiveness of storytelling on adult learning’, J. Workplace Learn., vol. 24, no. 6, pp. 430–438, Aug. 2012, doi: 10.1108/13665621211250333.	
\bibitem{b25}	H. B. Christensen, ‘A story-telling approach for a software engineering course design’, in Proceedings of the 14th annual ACM SIGCSE conference on Innovation and technology in computer science education, Paris France: ACM, Jul. 2009, pp. 60–64. doi: 10.1145/1562877.1562901.	
\bibitem{b26}	R. E. Landrum, K. Brakke, and M. A. McCarthy, ‘The pedagogical power of storytelling.’, Scholarsh. Teach. Learn. Psychol., vol. 5, no. 3, pp. 247–253, Sep. 2019, doi: 10.1037/stl0000152.	


\bibitem{b27}	G. M. Fernández-Nieto, V. Echeverria, R. Martinez-Maldonado, and S. B. Shum, ‘YarnSense: Automated Data Storytelling for Multimodal Learning Analytics’.	
\bibitem{b28}	P. Wuilmart, E. Söderberg, and M. Höst, ‘Programmer Stories, Stories for Programmers: Exploring Storytelling in Software Development’, in Companion Proceedings of the 7th International Conference on the Art, Science, and Engineering of Programming, Tokyo Japan: ACM, Mar. 2023, pp. 68–75. doi: 10.1145/3594671.3594677.	
\bibitem{b29}	J. Allen and C. Kelleher, ‘Exploring the impacts of semi-automated storytelling on programmers’ comprehension of software histories’, in 2024 IEEE Symposium on Visual Languages and Human-Centric Computing (VL/HCC), Liverpool, United Kingdom: IEEE, Sep. 2024, pp. 148–162. doi: 10.1109/VL/HCC60511.2024.00025.	
\bibitem{b30}	M. B. Kery, M. Radensky, M. Arya, B. E. John, and B. A. Myers, ‘The Story in the Notebook: Exploratory Data Science using a Literate Programming Tool’, in Proceedings of the 2018 CHI Conference on Human Factors in Computing Systems, Montreal QC Canada: ACM, Apr. 2018, pp. 1–11. doi: 10.1145/3173574.3173748.	



\bibitem{b31}	J. Allen, ‘Code Stories for Software Repurposing’, in 2023 IEEE Symposium on Visual Languages and Human-Centric Computing (VL/HCC), Washington, DC, USA: IEEE, Oct. 2023, pp. 309–311. doi: 10.1109/VL-HCC57772.2023.00063.	
\bibitem{b32}	M.-A. Storey, A. Zagalsky, F. F. Filho, L. Singer, and D. M. German, ‘How Social and Communication Channels Shape and Challenge a Participatory Culture in Software Development’, IEEE Trans. Softw. Eng., vol. 43, no. 2, pp. 185–204, Feb. 2017, doi: 10.1109/TSE.2016.2584053.	
\bibitem{b33}	M. Ito \textit{et al.}, ‘Connected learning: An agenda for research and design’, 2014. ISBN-13: 978-0-9887255-0-8



\bibitem{b34}	M. X. Liu, A. Kittur, and B. A. Myers, ‘Crystalline: Lowering the Cost for Developers to Collect and Organize Information for Decision Making’, in CHI Conference on Human Factors in Computing Systems, New Orleans LA USA: ACM, Apr. 2022, pp. 1–16. doi: 10.1145/3491102.3501968.	
\bibitem{b35}	A. Kharrufa, P. Olivier, and D. Leat, ‘Learning Through Reflection at the Tabletop: A Case Study with Digital Mysteries’, in Proceedings of EdMedia + Innovate Learning 2010, Waynesville, NC: Association for the Advancement of Computing in Education (, 2010, pp. 665–674.	
\bibitem{b36}	R. Chaudhury, T. Liaqat, and P. K. Chilana, ‘Exploring the Needs of Informal Learners of Computational Skills: Probe-Based Elicitation for the Design of Self-Monitoring Interventions’.	



\bibitem{b37}	A. Y. Wang, R. Mitts, P. J. Guo, and P. K. Chilana, ‘Mismatch of Expectations’, in Proceedings of the 2018 CHI Conference on Human Factors in Computing Systems, 2018. doi: 10.1145/3173574.3174085.	
\bibitem{b38}	M. Rõõm, M. Lepp, and P. Luik, ‘Dropout Time and Learners’ Performance in Computer Programming MOOCs’, Educ. Sci., vol. 11, 10, p. 643, 2021, doi: 10.3390/educsci11100643.	
\bibitem{b39}	A. Hawlitschek, V. Köppen, A. Dietrich, and S. Zug, ‘Drop-out in programming courses – prediction and prevention’, J. Appl. Res. High. Educ. Ahead--Print, 2019, doi: 10.1108/jarhe-02-2019-0035.	
\bibitem{b40}	L. E. Margulieux, B. B. Morrison, and A. Decker, ‘Reducing withdrawal and failure rates in introductory programming with subgoal labeled worked examples’, Int. J. STEM Educ., vol. 7, no. 1, p. 19, Dec. 2020, doi: 10.1186/s40594-020-00222-7.	



\bibitem{b41}	D. F. O. Onah and J. E. Sinclair, ‘Assessing Self-Regulation of Learning Dimensions in a Stand-alone MOOC Platform’, Int. J. Eng. Pedagogy IJEP, vol. 7, no. 2, p. 4, May 2017, doi: 10.3991/ijep.v7i2.6511.	
\bibitem{b42}	G. Schraw, K. J. Crippen, and K. Hartley, ‘Promoting Self-Regulation in Science Education: Metacognition as Part of a Broader Perspective on Learning’, Res. Sci. Educ., vol. 36, no. 1, pp. 111–139, Mar. 2006, doi: 10.1007/s11165-005-3917-8.	



\bibitem{b43}	L. Silva, A. Mendes, A. Gomes, and G. Fortes, ‘What Learning Strategies are Used by Programming Students? A Qualitative Study Grounded on the Self-regulation of Learning Theory’, ACM Trans. Comput. Educ., vol. 24, no. 1, pp. 1–26, Mar. 2024, doi: 10.1145/3635720.	
\bibitem{b44}	B. J. Zimmerman, ‘Attaining self-regulation: A social cognitive perspective’, in Handbook of self-regulation, Elsevier, 2000, pp. 13–39.	
\bibitem{b45}	G. Ciolacu, C. Haas, and M. Hall, ‘Rightsizing: Understanding Novice, Casual Learners of Programming’, in Proceedings of the 56th ACM Technical Symposium on Computer Science Education V. 2, Pittsburgh PA USA: ACM, Feb. 2025, pp. 1421–1422. doi: \url{10.1145/3641555.3705239}.	


\bibitem{b46}	A. B. Morrison and L. L. Richmond, ‘Offloading items from memory: individual differences in cognitive offloading in a short-term memory task’, Cogn. Res. Princ. Implic., vol. 5, no. 1, p. 1, Dec. 2020, doi: 10.1186/s41235-019-0201-4.	
\bibitem{b47}	S. J. Gilbert, A. Boldt, C. Sachdeva, C. Scarampi, and P.-C. Tsai, ‘Outsourcing Memory to External Tools: A Review of “Intention Offloading”’, Psychon. Bull. Rev., vol. 30, no. 1, pp. 60–76, Feb. 2023, doi: 10.3758/s13423-022-02139-4.	
\bibitem{b48}	H. S. Meyerhoff, S. Grinschgl, F. Papenmeier, and S. J. Gilbert, ‘Individual differences in cognitive offloading: a comparison of intention offloading, pattern copy, and short-term memory capacity’, Cogn. Res. Princ. Implic., vol. 6, no. 1, p. 34, Dec. 2021, doi: 10.1186/s41235-021-00298-x.	

\bibitem{b49}	P. K. Chilana et al., ‘Perceptions of non-CS majors in intro programming: The rise of the conversational programmer’, in 2015 IEEE Symposium on Visual Languages and Human-Centric Computing (VL/HCC), Atlanta, GA: IEEE, Oct. 2015, pp. 251–259. doi: \url{10.1109/VLHCC.2015.7357224}.	


\bibitem{b50}	K. R. Vareberg and C. A. Platt, ‘Harnessing the wisdom of YouTube: how self-directed learners achieve personalized learning through technological affordances’, Interact. Learn. Environ., vol. 32, no. 10, pp. 7141–7155, Nov. 2024, doi: 10.1080/10494820.2024.2307597.	
\bibitem{b51}	D. Lambton-Howard, J. Kiaer, and A. Kharrufa, ‘“Social media is their space”: student and teacher use and perception of features of social media in language education’, Behav. Inf. Technol., vol. 40, no. 16, pp. 1700–1715, Dec. 2021, doi: 10.1080/0144929X.2020.1774653.	
\bibitem{b52}	N. Dabbagh and A. Kitsantas, ‘Personal Learning Environments, social media, and self-regulated learning: A natural formula for connecting formal and informal learning’, Internet High. Educ., vol. 15, no. 1, pp. 3–8, Jan. 2012, doi: 10.1016/j.iheduc.2011.06.002.


\bibitem{b53}
T.~Faas, L.~Dombrowski, E.~Brady, and A.~Miller,
“Looking for Group: Live Streaming Programming for Small Audiences,”
in *Information in Contemporary Society*, 2019, pp.~117--123.
doi: \url{10.1007/978-3-030-15742-5_10}.




\bibitem{b54}	R. M. Ryan and E. L. Deci, ‘Self-determination theory and the facilitation of intrinsic motivation, social development, and well-being’, Am. Psychol., vol. 55, 1, pp. 68–78, 2000, doi: 10.1037/0003-066x.55.1.68.	
\bibitem{b55}	Z. Duo, J. Zhang, Y. Ren, and X. Xu, ‘Examining self-regulation models of programming students in visual environments: A bottom-up analysis of learning behaviour’, Educ. Inf. Technol., vol. 30, no. 4, pp. 5229–5249, Mar. 2025, doi: 10.1007/s10639-024-13016-z.	
\bibitem{b56}	P. Gu, J. Wu, Z. Cheng, Y. Xia, M. Cheng, and Y. Dong, ‘Scaffolding self-regulation in project-based programming learning through online collaborative diaries to promote computational thinking’, Educ. Inf. Technol., Feb. 2025, doi: 10.1007/s10639-025-13367-1.	
\bibitem{b57}	G. Cheng, D. Zou, H. Xie, and F. L. Wang, ‘Exploring differences in self-regulated learning strategy use between high- and low-performing students in introductory programming: An analysis of eye-tracking and retrospective think-aloud data from program comprehension’, Comput. Educ., vol. 208, p. 104948, Jan. 2024, doi: 10.1016/j.compedu.2023.104948.	


\bibitem{b58}	K. Falkner, R. Vivian, and N. J. G. Falkner, ‘Identifying computer science self-regulated learning strategies’, in Proceedings of the 2014 conference on Innovation \& technology in computer science education - ITiCSE ’14, Uppsala, Sweden: ACM Press, 2014, pp. 291–296. doi: 10.1145/2591708.2591715.	


\bibitem{b59}	D. J. Ferreira, D. S. Campos, and A. C. Gonçalves, ‘Regulatory Strategies for Novice Programming Students’, in Computer Supported Education, vol. 2052, B. M. McLaren, J. Uhomoibhi, J. Jovanovic, and I.-A. Chounta, Eds., in Communications in Computer and Information Science, vol. 2052. , Cham: Springer Nature Switzerland, 2024, pp. 136–159. doi: \url{10.1007/978-3-031-53656-4_7}.

\bibitem{b60}	R. Pérez-Álvarez, J. Maldonado-Mahauad, and M. Pérez-Sanagustín, ‘Tools to Support Self-Regulated Learning in Online Environments: Literature Review’, in Lifelong Technology-Enhanced Learning, vol. 11082, V. Pammer-Schindler, M. Pérez-Sanagustín, H. Drachsler, R. Elferink, and M. Scheffel, Eds., in Lecture Notes in Computer Science, vol. 11082. , Cham: Springer International Publishing, 2018, pp. 16–30. doi:\url{10.1007/978-3-319-98572-5_2}.	

\bibitem{b61}	J. Broadbent, E. Panadero, J. M. Lodge, and P. Barba, ‘Technologies to Enhance Self-Regulated Learning in Online and Computer-Mediated Learning Environments’, Handb. Res. Educ. Commun. Technol., pp. 37–52, 2020, doi: \url{10.1007/978-3-030-36119-8_3}.	

\bibitem{b62}	R. Chaudhury, ‘Designing Interactive Self-Monitoring Tools for Informal Learners of Computational Skills’, in 2023 IEEE Symposium on Visual Languages and Human-Centric Computing (VL/HCC), Washington, DC, USA: IEEE, Oct. 2023, pp. 307–308. doi: \url{10.1109/VL-HCC57772.2023.00062}.	


\bibitem{b63}	U. Lyngs et al., ‘Self-Control in Cyberspace: Applying Dual Systems Theory to a Review of Digital Self-Control Tools’, in Proceedings of the 2019 CHI Conference on Human Factors in Computing Systems, Glasgow Scotland Uk: ACM, May 2019, pp. 1–18. doi: 10.1145/3290605.3300361.	
\bibitem{b64}	A. M. Roffarello and L. De Russis, ‘Achieving Digital Wellbeing Through Digital Self-control Tools: A Systematic Review and Meta-analysis’, ACM Trans. Comput.-Hum. Interact., vol. 30, no. 4, pp. 1–66, Aug. 2023, doi: 10.1145/3571810.	
\bibitem{b65}	R. Wang, A. Abusafia, A. Lakhdari, and A. Bouguettaya, ‘The Nudging Effect on Tracking Activity’, in Proceedings of the 2022 ACM International Joint Conference on Pervasive and Ubiquitous Computing, Cambridge United Kingdom: ACM, Sep. 2022, pp. 130–132. doi: 10.1145/3544793.3560366.	
\bibitem{b66}	Forest: Stay focused, be present. Accessed: May 03, 2025. [Online]. Available: https://www.forestapp.cc/	
\bibitem{b67}	Habitica: Gamify Your Life. Accessed: May 03, 2025. [Online]. Available: https://habitica.com/static/home	
\bibitem{b68}	C. Wang, ‘Comprehensively Summarizing What Distracts Students from Online Learning: A Literature Review’, Hum. Behav. Emerg. Technol., vol. 2022, pp. 1–15, Oct. 2022, doi: 10.1155/2022/1483531.	
\bibitem{b69}	C. Yot-Domínguez and C. Marcelo, ‘University students’ self-regulated learning using digital technologies’, Int. J. Educ. Technol. High. Educ., vol. 14, no. 1, p. 38, Dec. 2017, doi: 10.1186/s41239-017-0076-8.	
\bibitem{b70}	W. J. Joel, ‘Engaging computer science education’, ACM SIGCSE Bull., vol. 38, no. 3, pp. 316–316, Sep. 2006, doi: 10.1145/1140123.1140222.	

\bibitem{b71}	W. J. Joel, ‘A story paradigm for computer science education’, in Proceedings of the 18th ACM conference on Innovation and technology in computer science education, Canterbury England, UK: ACM, Jul. 2013, pp. 362–362. doi: 10.1145/2462476.2466526.	


\bibitem{b72}	A. Korhonen and M. Vivitsou, ‘Digital Storytelling and Group Work: Integrating the Narrative Approach into a Higher Education Computer Science Course’, in Proceedings of the 2019 ACM Conference on Innovation and Technology in Computer Science Education, Aberdeen Scotland Uk: ACM, Jul. 2019, pp. 140–146. doi: 10.1145/3304221.3325528.	


\bibitem{b73}	A. Horvath, ‘Meta-Information to Support Sensemaking by Developers’.	
\bibitem{b74}	A. Horvath, A. Macvean, and B. A. Myers, ‘Support for Long-Form Documentation Authoring and Maintenance’, in 2023 IEEE Symposium on Visual Languages and Human-Centric Computing (VL/HCC), Washington, DC, USA: IEEE, Oct. 2023, pp. 109–114. doi: 10.1109/VL-HCC57772.2023.00020.	


\bibitem{b75}	A. Horvath, B. Myers, A. Macvean, and I. Rahman, ‘Using Annotations for Sensemaking About Code’, in Proceedings of the 35th Annual ACM Symposium on User Interface Software and Technology, Bend OR USA: ACM, Oct. 2022, pp. 1–16. doi: 10.1145/3526113.3545667.	


\bibitem{b76}	L. Margulieux and R. Catrambone, ‘Using Learners’ Self-Explanations of Subgoals to Guide Initial Problem Solving in App Inventor’, in Proceedings of the 2017 ACM Conference on International Computing Education Research, Tacoma Washington USA: ACM, Aug. 2017, pp. 21–29. doi: 10.1145/3105726.3106168.	


\bibitem{b77}	A. Horvath et al., ‘Understanding How Programmers Can Use Annotations on Documentation’, in CHI Conference on Human Factors in Computing Systems, New Orleans LA USA: ACM, Apr. 2022, pp. 1–16. doi: 10.1145/3491102.3502095.	


\bibitem{b78}	D. A. Kolb, Experiential Learning. Pearson Education, 2014.	


\bibitem{b79}	New General Self-Efficacy Scale. [Online]. Available: \url{https://sparqtools.org/mobility-measure/new-general-self-efficacy-scale/}	

\bibitem{b80}	N. Uzdil and Y. Günaydın, ‘The effect of sense of coherence on mindful attention awareness and academic self-efficacy in nursing students’, Nurse Educ. Pract., vol. 64, p. 103429, Oct. 2022, doi: 10.1016/j.nepr.2022.103429.	
\bibitem{b81}	K. Konaszewski, M. Kolemba, and M. Niesiobędzka, ‘Resilience, sense of coherence and self-efficacy as predictors of stress coping style among university students’, Curr. Psychol., vol. 40, no. 8, pp. 4052–4062, Aug. 2021, doi:\url{10.1007/s12144-019-00363-1}.	



\bibitem{b82}	A. Masry Herzallah and R. Makaldy, ‘Technological self-efficacy and sense of coherence: Key drivers in teachers’ AI acceptance and adoption’, Comput. Educ. Artif. Intell., vol. 8, p. 100377, Jun. 2025, doi: 10.1016/j.caeai.2025.100377.	
\bibitem{b83}	Y. Salamonson, L. M. Ramjan, S. Van Den Nieuwenhuizen, L. Metcalfe, S. Chang, and B. Everett, ‘Sense of coherence, self-regulated learning and academic performance in first year nursing students: A cluster analysis approach’, Nurse Educ. Pract., vol. 17, pp. 208–213, Mar. 2016, doi: 10.1016/j.nepr.2016.01.001.	
\bibitem{b84}	‘UEQ-S’, User Experience Questionnaire. [Online]. Available: https://www.ueq-online.org/	


\bibitem{zimmerman2007}
J.~Zimmerman, J.~Forlizzi, and S.~Evenson, ``Research through design as a method for interaction design research in {HCI},'' in \emph{Proceedings of the SIGCHI Conference on Human Factors in Computing Systems}, San Jose, California, USA: ACM, Apr. 2007, pp. 493--502. doi: \url{10.1145/1240624.1240704}.

\bibitem{wickedproblem}
``\emph{What’s a Wicked Problem?}'' [Online]. Available: \url{https://www.stonybrook.edu/commcms/wicked-problem/about/What-is-a-wicked-problem}

\bibitem{lippke2020}
S.~Lippke, ``Self-efficacy theory,'' in \emph{Encyclopedia of Personality and Individual Differences}, V.~Zeigler-Hill and T.~K.~Shackelford, Eds. Cham: Springer, 2020, pp.~4722--4727. doi: 10.1007/978-3-319-24612-3\_1167.


\bibitem{prolific}
Prolific, ``Prolific: Participant recruitment for research,'' [Online]. Available: \url{https://www.prolific.com/}

\bibitem{pajares1996}
F.~Pajares, ``Self-efficacy beliefs in academic settings,'' \emph{Rev. Educ. Res.}, vol.~66, no.~4, pp.~543--578, 1996, doi: 10.2307/1170653.


\end{thebibliography}
\end{document}